# First On-Site True Gamma-Ray Imaging-Spectroscopy of Contamination near Fukushima Plant

Dai Tomono[1], Tetsuya Mizumoto[1], Atsushi Takada[1], Shotaro Komura[1], Yoshihiro Matsuoka[1], Yoshitaka Mizumura[1,2], Makoto Oda[1], and Toru Tanimori[1,2*]

1 Graduate School of Science, Kyoto University, Sakyo, Kyoto, 606-8502, Japan

2 Unit of Synergetic Studies for Space, Kyoto University, Sakyo, Kyoto, 606-8502, Japan

**We have developed an Electron Tracking Compton Camera (ETCC), which provides a well-defined Point Spread Function (PSF) by reconstructing a direction of each gamma as a point and realizes simultaneous measurement of brightness and spectrum of MeV gamma-rays for the first time.**

**Here, we present the results of our on-site pilot gamma-imaging-spectroscopy with ETCC at three contaminated locations in the vicinity of the Fukushima Daiichi Nuclear Power Plants in Japan in 2014. The obtained distribution of brightness (or emissivity) with remote-sensing observations is unambiguously converted into the dose distribution. We confirm that the dose distribution is consistent with the one taken by conventional mapping measurements with a dosimeter physically placed at each grid point. Furthermore, its imaging spectroscopy, boosted by Compton-edge-free spectra, reveals complex radioactive features in a quantitative manner around each individual target point in the background-dominated environment. Notably, we successfully identify a "micro hot spot" of residual caesium contamination even in an already decontaminated area. These results show that the ETCC performs exactly as the geometrical optics predicts, demonstrates its versatility in the field radiation measurement, and reveals potentials for application in many fields, including the nuclear industry, medical field, and astronomy.**

**Introduction**

Following the accident in Fukushima Daiichi Nuclear Power Plants on 11 March 2011, a huge amount of radionuclides was released to the atmosphere. As in 2016, $^{137}$Cs and $^{134}$Cs, which radiate gammas mainly from 600 keV to 800 keV, still remain in Fukushima, and many areas are still contaminated as a result[1]. Operations of decontamination are called for in a wide area in Fukushima and its surroundings to satisfy a legal limit for the maximum exposure of 0.23 μSv/h at any publicly-accessible open spaces[2]. An effective method to measure and monitor gamma-ray radiation is essential for efficient decontamination work, and as a result there has been a surge of demand for gamma-ray instruments with a wide field of view (FoV) which quantitatively visualize Cs contamination.

Many gamma cameras have been developed to make imaging observations to help decontamination, based on the Compton camera (CC)[3-7], pin-hole (PHC)[8], and coded-mask technologies. However, none of them has detected more than a limited number of hot spots, or has reported any quantitative radiation maps, let alone imaging spectroscopy. The CC is the most advanced among these three, yet has an intrinsic difficulty in imaging spectroscopy, which is related to its Point Spread Function (PSF)[9, 10].

So far, the most successful evaluations for the environmental radiation in contamination areas have been made by backpacks[11] and unmanned helicopters[12, 13]. Although these methods are, unlike gamma cameras, non-imaging measurements, in which measurements at each point are made with either a spectrometer or conventional dosimeter, quantitative and reliable 2-dimensional distributions of radiation have been successfully obtained after several measurements with overlapping fields of view are combined. The downside is that they require a considerable amount of time and efforts, and thus are not practical to be employed in a wide area.

Another fundamental problem with all these methods is that they do not directly measure the radioactivity on the ground, but measure the dose at 1 metre high from the ground (hereafter referred to as "1-m dose") instead, and hence require complex analyses to convert the measured dose to the actual radioactivity on the ground. Indeed, we show that the 1-m dose does not always agree well with that measured immediately above the ground, which suggests an intrinsic difficulty in obtaining an accurate radioactivity distribution on the ground from the 1-m dose.

After a few pilot experiments of decontamination were conducted in Fukushima, it turned out that the amount of reduction of the ambient dose by decontamination was limited. The reduction ratios, defined by the dose ratio compared between before and after decontamination, were approximately 20 % only in lower ambient-dose areas (<3 μSv/h) [2], while >39 % in higher ambient-dose areas (>3 μSv/h). When a (high) dose is measured at a point, gammas that contribute to the dose can originate anywhere a few radiation lengths away (~100 m) from the point. The goal of decontamination is to somehow identify and remove those radiation sources. However, none of the existing instruments can identify them, i.e., none of them can tell where or even in which direction the radiation source is located. To untangle the sources of a dose of contamination, the directions of all the gammas, as well as their energies if possible, must be determined. It means that the brightness distribution around the point must be obtained.

To address these issues of existing methods and visualize the Cs contamination, we have developed and employed an Electron-Tracking Compton Camera (ETCC). ETCCs were originally developed to observe nuclear gammas from celestial objects in MeV astronomy[14], but have been applied in wider fields, including medical imaging[15] and environmental monitoring[16,17]. An ETCC outputs two angles of an incident gamma by measuring the direction of a recoil electron and hence provides the brightness distribution of gammas with a resolution of the PSF[9,10]. The PSF is determined from the angular resolutions of angular resolution measure (ARM) and scatter plane deviation (SPD)[9, 18]. The ARM and SPD correspond to a resolution of the polar and azimuthal angles of an incident gamma, respectively. Since a leakage of gammas from their adjacent region to the measured point is correctly estimated with the PSF, quantitative evaluation of the emissivity anywhere in the FoV is attained.

The most remarkable feature of the ETCC is to resolve the Compton process completely; the ETCC does not only provide the direction of a gamma, but also enables us to distinguish correctly reconstructed gammas from those mis-reconstructed[9]. Thus, the ETCC makes true images of gammas based on proper geometrical optics (PGO), as well as energy spectra[9] free of Compton edges[10]. The PGO enables us to measure precise brightness (or emissivity) at any points in an image using an equi-solid-angle projection, such as Lambert projection, without the information of the distance to the source, as shown in Fig.1. The obtained emissivity can be unambiguously converted into the dose on the ground (hereafter the E-dose), of which the procedure is identical with that described in the IAEA report[19], but without need of the fitting parameters. We find the E-dose to be consistent with the dose independently measured by a dosimeter, and thus confirm that remote-sensing imaging-spectroscopy with the ETCC perfectly reproduces the spatial distribution of radioactivity[10].

**Results**

We performed the field test of gamma measurement in October, 2014 in relatively high-dose locations with the averaged ambient dose ranging from 1 to 5 μSv/h in Fukushima prefecture, using the compact 10 cm x 10 cm x 16 cm ETCC with a FoV of ~100°ϕ[17]. The SPD and ARM of the ETCC were measured to be 120° and 6° (FWHM), respectively, for 662-keV $^{137}$Cs peaks, which correspond to the PSF (Θ~15°), i.e., the radius of the PSF of 15° for the region that encompasses a half of gammas emitted from a point source[9]. It uses GSO scintillators and has an energy resolution of 11% (FWHM) at 662 keV. We chose the three different kinds of locations for measurements: (A) decontaminated pavement surrounded with not-decontaminated bush, (B) not-decontaminated ground, and (C) decontaminated parking lot. Figs.2a, 2c, and 2d show their respective photographs.

We have found that the doses at 1-m and 1-cm measured with a dosimeter do not agree with each other, as demonstrated in Figs. 2a and 2b in the location (C). The 1-m dose, which is practically the emissivity averaged over the adjacent region of ~10 m, is the standard in the radiation measurement, presumably because it is useful to estimate potential health effects to the human body. The 1-cm dose, on the other hand, better reflects the emissivity on the ground at each grid point, of which the size is likely to be similar to the spatial resolution of the ETCC, and hence is useful to locate radioactivity on the ground for decontamination work. For these reasons, we adopt the 1-cm dose to compare with the emissivity measured with the ETCC in this work.

Figure 3a shows the photograph of FoV, overlaid with 1-cm dose at nine points and the E-dose map by ETCC, where the brightness (equivalent to the E-dose) is defined as the count rate of reconstructed gammas per unit solid angle (here 0.014 sr), corrected for the detection efficiency including the angular dependence of the ETCC[9]. Fig.3b shows the energy spectrum accumulated for the entire FoV, whereas Figs.3c-3e display those accumulated for the sky, the decontaminated pavement, and the not-decontaminated bush, respectively. The E-dose at the maximum brightness in Fig.3a is estimated to be 2.6 μSv/h, which is consistent with the average of the 1-cm dose (0.9-4.3 μSv/h) around the centre of the FoV.

The spatial distribution of the E-dose is found to be consistent with that of the 1-cm dose, which was independently measured. The spectrum in Fig.3e shows prominent peaks of direct gammas of Cs, which implies the contamination from the bush area, whereas the spectrum of the decontaminated pavement (Fig.3d) shows much weaker Cs peaks, which implies the effect of the decontamination. The latter is dominated with low-energy scattered gammas, which emanate from inside of the ground and the adjacent areas. The spectrum of the sky (Fig.3c) is clearly dominated with Compton-scattered gammas from Cs peaks (with the expected energy ranging from 200 to 500 keV) in the air. We should note that the spectra free of Compton scattering components enable us to make the unambiguous identification of the sources of radiation.

The results of imaging-spectroscopy in the two contrasting locations (B and C), in which no and thorough decontaminations, respectively, have been conducted, are shown in Fig.4 and Fig.5. The exposure times are 80 min and 100 min, respectively. The ETCC gives spatially-resolved spectra, and accordingly the detailed condition of contamination at each point, similar to Fig.3. In the contaminated location (B), although the energy spectrum of the FoV shows strong and direct gamma emission from Cs, Cs is found to be concentrated in the limited area of spot 1 (Fig.4e), whereas little Cs is found in the other regions in the FoV (Fig.4f). As such, imaging-spectroscopic measurement is a reliable method to unravel the state of contamination quantitatively. Even in the decontaminated location (C), both the image (Fig.5a) and spectrum (Fig.5f) reveal the existence of a "micro hot spot", where some Cs remains on the ground and the spectrum has the dominant Cs peak (Fig.5f), whereas the spectra for other regions (Fig.5e) show that the main component is scattered low-energy gammas. Both the maps of 1-cm dose (Fig.5a) and E-dose (Fig.5b) show a hint of a small enhancement originating from a micro hot spot, although it is at a similar level to the fluctuation of the scattered gammas. The E-doses at the points of the maximum brightness in (B) and (C) are 5.0 and 1.3 μSv/h, respectively, which are also consistent with the 1-cm dose at the corresponding points.

Finally, we check consistency about a couple of properties of the ETCC and conventional dose measurements. First, we plot the total gamma counts obtained with the ETCC as 1-m doses at the position of the ETCC in Fig.6a, and confirm a good correlation. Then, we plot the correlation between the 1-cm dose measured by the dosimeter and by the ETCC (E-dose) at the locations (B) and (C) in Fig.6b. Except ~3 points adjacent to the hot spots in (B), the discrepancy between them is limited within ±~30 %. Considering the difference in the conditions, such as the size of the measured areas

(~100 cm² for a dosimeter and ~1 m² for ETCC) and the energy range (>150 keV for a dosimeter and 486-1000 keV for ETCC), as well as the fact that a large dispersion in the accuracy of commercial dosimeters (± several 10 %) has been reported, this amount of discrepancy is more or less expected. We conclude that good consistency between them is established for the wide range of the dose (0.1-5 μSv/h), and this is another proof that the ETCC achieves the PGO. In addition, the PGO gives the brightness of the sky over the hemisphere, and we find it to be comparable with that from the ground, after the difference in their solid angles is corrected (see the bottom row in Table 1). This means that roughly a half of the 1-m dose at any points originates from the sky. It then implies that the wide-band energy balance of gammas between the ground and the sky is in equilibrium and contribute to the ambient dose, presumably because the air is thick enough to scatter most of gammas emanating from the ground. It is consistent with the fact that the spectra of the sky (Fig.3c, Fig.4c, Fig.5c) are dominated with Compton scattering for Cs gammas (200-500 keV). This could not have been identified without spectra free of Compton edges. Our results also explain the reason why the amount of the reduction of the ambient dose was limited to often no more than 50% after decontamination work[2] had been conducted in Fukushima, it is because a significant amount of radiation still comes from the sky in equilibrium.

**Discussion**

Firstly, let us convert the emissivity to the 1-cm dose, using only the brightness measured by the ETCC. Figure 1b schematically shows the dosimeter configuration for the measurement of 1-cm dose. Since the top and the upper sides of the dosimeter are shielded with tungsten (W) rubber, it detects gammas emanating from the ground to the lower hemisphere only. The count density of the gammas which pass through the plane of the dosimeter (indicated as P in Fig.1b) is estimated to be approximately $2\pi\Sigma \cdot (1 - \cos(\theta=80°)) = 5.2\Sigma$, where $\Sigma$ is emissivity on the ground. Then we convert the count density of gammas at the dosimeter position into doses in units of μSv/h with the conversion factor of 1 μSv/h = ~100 counts・sec$^{-1}$・cm$^{-2}$ for 662-keV gammas in the dosimeter, based on the IAEA report[19] (in page 85).

In the not-decontaminated location (B), 135 gammas were observed with the ETCC (*dB*) at the maximum brightness point in Fig.4a, where the unit solid angle is 0.014 sr. The brightness of the gamma is calculated to be 135 count / ( sec・.014 sr・100 cm² ) = 96 counts・sec$^{-1}$・sr$^{-1}$・cm$^{-2}$, and then we get, from the relation $\Sigma = dB$, $5.2\Sigma = 500$ counts・sec$^{-1}$・cm$^{-2}$, which corresponds to the dose of 5.0 μSv/h (the two points indicated as 5.0 and 5.7 [μSv/h] in Fig.4a). For the location (C), 35 gammas were observed at the maximum brightness point in Fig. 5a, and then $dB (= \Sigma) = 35$ counts / (sec・0.014 sr・100 cm²) = 25 counts・sec$^{-1}$・sr$^{-1}$・cm$^{-2}$ and $5.2\Sigma = 130$ counts sec$^{-1}$・cm$^{-2}$, which corresponds to 1.3 μSv/h. The 1-cm dose at this point is found to be roughly equal to the average of 1.0 - 2.2 μSv/h in Fig.5a. For the location (A), *dB* is calculated in the similar manner to be *dB* = 70 counts / ( sec・0.014 sr・100 cm$^2$) = 50 counts・sec$^{-1}$・sr$^{-1}$・cm$^{-2}$ and 5.2$\Sigma$ = 260 counts・sec$^{-1}$・cm$^{-2}$, which corresponds to the dose of 2.6μSv/h. The 1-cm dose at this point is ~3 μSv/h, and is roughly equal to the average of 1 - 4.3 μSv/h in Fig.3a.

For comparison, we also applied the simple method described in pages 96-101 in the IAEA report[19], calculating the doses with a conversion coefficient of 8.7 x 10$^{-3}$ (μSv/h) / (Bq/cm$^2$) for θ~80 º for the 1-cm dose, which is estimated by accumulating gamma-flux at each point from the ground with the tungsten rubber shield. This method is the one described in pages 96-101 in the IAEA report[19]. For the location (A), a gamma flux on the ground is calculated to be 2π$\Sigma$/0.85=369 (Bq/cm$^2$) and then the dose is 369 x 8.7 x 10$^{-3}$ = 3.1 μSv/h. For the locations (B) and (C), the doses are estimated to be 5.9 and 1.6 μSv/h, respectively. Thus, we confirmed that the results deduced by the two independent methods are consistent with each other.

Decontamination work in Fukushima faces serious difficulty; it is hard to pin down which region is badly contaminated from which radiation source without investing massive resources like wide-scale backpack measurements. The capability of the ETCC to measure the emissivity (or dose) independently of the distance would enable us to propose a novel approach to it. If a mapping of the brightness of $^{137}$Cs on the ground was carried out over the wide area with the ETCC by aircraft with the similar way conducted in 2012[2], we could visualize variation of the doses across the area, and could tell where decontamination work would be required most and how much.

As a different application, if multiple ETCCs are installed at various places in a nuclear plant to carry out a continuous three-dimensional brightness monitoring, we could not only detect, for example, a sudden radiation release by accident, but also make a quantitative assessment of where and how the release has happened. This would provide vital initial parameters to computer simulations to estimate the later dissemination of radioactivity over a wide area after an accident. In fact, simulations for this purpose faced a great difficulty in the past due to lack of reliable observed parameters of radio activity, because radiation monitoring was performed solely by repeated simple dose measurements. These simple dose measurements are unable to provide sufficient information over the wide area where the gamma radiation comes from, unless a huge amount of resources of manpower and hence budget are invested. Given that governments in many countries are confronted with the reactor dismantling issue, detailed and quantitative mapping of the radiation emissivity on the surfaces of reactor facilities, which would be well achievable with the ETCC, would be beneficial. The ETCC has immense potentials for immediate applications to various radiation-related issues in the environment.

**Prospects**

Some scientists assert that the detection efficiency of gas-based gamma detectors would be too low. However, we have found that some types of gas have sufficient Compton-scattering probability with the relevant effective areas of 110 cm$^2$ and 65 cm$^2$ at 1-MeV gammas with a 50-cm-cubic ETCC using

$CF_4$ gas and Ar gas at 3 atm, respectively[9]. Our prototype 30 cm-cubic ETCC with the effective area of a few $cm^2$ at 300 keV was proved to perform expectedly well in MeV gamma-ray astronomy.

Now, we are constructing two types of more advanced ETCCs: one is a compact ETCC with the similar size and weight to the current model, but having a 20 times larger effective area (0.2 $cm^2$ at 662 keV; type-A) and the other is a large ETCC aimed to be completed in 2018, which has a 1000 times larger effective area (10 $cm^2$ at 662 keV; type-B). The details of Type-B are described elsewhere[10].

Type-A has the similar size to the current ETCC, but has an increased TPC volume from 10 cm x 10 cm x 16 cm (rectangular solid) to 20 cm$\phi$ (in diameter) x 20 cm (cylinder), installed in the similar-sized gas vessel. It has a 5 times larger gas volume and 2.5 times wider detectable electron energy band with the TPC than the current model. In addition, if the mixed gas with Ar and $CF_4$ (50 % : 50 %) at 2 atm is used, as opposed to the current Ar gas (~90% and some cooling gases) at 1.5 atm, the detection efficiency will be improved by a factor of $2^9$. Then, the resultant detection efficiency (or effective area) will become 20 times larger than that of the current model, while keeping the similarly compact size and weight. The development of Type-A will be completed in 2017.

Type-B will provide the same detection limit for 6 sec exposure. If we perform a survey with Type-B from some aircraft at the altitude of 100 m, we will be able to make a spectroscopic map of a 1 $km^2$ area with a 10 m x 10 m resolution for 1200 sec exposure to achieve the same detection limit, taking account of the absorption of the air. An unmanned airship is a good candidate for the aircraft, it flies slowly for an extended period and hence would enable us to do the precise imaging-spectroscopic survey. Then, the whole contamination area in Fukushima prefecture (roughly 20 km x 50 km) can be mapped with the same resolution as mentioned above in a realistic timescale of~2 months, assuming 8 hours of work per day. Some of the spectra obtained in our aircraft-based survey might be found out to be generated by the gammas scattered by something, such as trees in woods, within the grid. Our survey will efficiently detect a hint for those areas, which can be then studied in more detail with on-site measurements, such as ones by backpacks[11]. No successful large-scale survey has been yet performed to monitor the radioactivity in Fukushima. Our upgraded ETCC will be capable of revolutionizing the decontamination work and more. We summarized the specifications of the current ETCC, type-A and type-B in Table 2.

**Methods**
**Instruments and Measurements**

The ETCC was mounted at 1.3 m high from the ground at its centre, tilted 20º downwards beneath the horizontal plane. The average distance to the ground in the FoV is ~4 m, which corresponds to the spatial resolution of ~1 m at the ground for its PSF. As a reference, we also made a mapping measurement of the dose at two heights of 1 m and 1 cm with every 1-m square grid in the FoV (except for the location (A), where the points of the measurements were sparser and irregular) with the commercial dosimeter (HORIBA, Radi PA-1100, http://www.horiba.com). In the dose measurement at the latter height (~1 cm), the top and four sides of the dosimeter were covered by tungsten rubber to shield it from the downward radiation (Fig.1b).

We have developed a compact ETCC with a 10 cm x 10 cm x 16 cm gas volume, based on the 30-cm-cubic SMILE-II for MeV astronomy[9]. The ETCC is, like CCs, equipped with a forward detector as a scatterer of nuclear gammas and a backward detector as a calorimeter for measuring the energy and hit position of scattered gammas. The forward detector of the ETCC is a gaseous Time Projection Chamber (TPC) based on micro-pattern gas detectors (MPGD), which tracks recoil electrons. The TPC of the ETCC is a closed gas chamber, and thus can be used continuously for about three weeks without refilling with the gas[5]. The backward detector is pixel scintillator arrays (PSAs) with heavy crystal (at present we use $Gd_2SiO_5$:Ce, GSO). It is noted that, at the time of writing in 2016 after the survey work presented in this paper, we have been developing the Ethernet-based data handling system to replace the existing VME-based system. The latest ETCC available for field measurements is much more compact, which is built in the 40 cm x 40 cm x 50 cm base frame with the weight of 40-50 kg, and operated with a single PC with 24 V portable battery.

The contamination area in Fukushima is the similar environment to the space in the background dominated condition, where the radiation spreads ubiquitously. It is understandable that gamma cameras with the Compton method became the first choice to be employed for the decontamination work in Fukushima, following the precedents in MeV astronomy, even though it is clearly not the ideal instrument especially in the background-dominated environment.

**Analytical method for deriving an emissivity from the measured distribution of gammas**

Here, we explain how we measure the emissivity (or brightness) based on the proper geometrical optics (PGO) by the ETCC and how we estimate the dose on the ground from the emissivity measured by the ETCC. The following are the reason why no gamma camera but the ETCC can take a quantitative nuclear gamma image with the similar principle to that of optical cameras. According to the well-known formulas in PGO, the relation between emissivity $\Sigma$ on the ground and detected brightness of the gamma in ETCC ($dB$) for solid angle $\Omega$ is given as $\Sigma \cdot A_1 \cdot d\Omega_1 = dB \cdot A_2 \cdot d\Omega_2$, and the relations $d\Omega_1 = A_2/D^2$, $d\Omega_2 = A_1/D^2$ hold, where $A_1$ and $A_2$ are the observed areas on the ground and the detection area in the ETCC ($A_2 = 100$ cm$^2$), respectively, and $D$ is a distance between the ground and the ETCC. Fig.1a gives a schematic demonstration of it. These relations are then reduced to $\Sigma = dB$, which means that the emissivity is equal to the obtained brightness and is independent of the distance $D$ in this optics. In practice, $dB$ is calculated simply from the number of the detected gammas per unit solid angle corrected for the detection efficiency[9]. We should note that when the distance between a source and the ETCC (*L*) is comparable with, or longer than, the radiation length in the air (~70 m), *dB* in a unit solid angle must be corrected for the expected absorption, using the absorption coefficient ($\alpha$) in the air for gammas with the relation $dB_{correct} = dB / (1 - exp(-L/\alpha))$.

**Estimation of the emissivity and the detection limits**

We estimate the detection limit using the sensitivity from the calibration data with a point source ($^{137}$Cs, 3 MBq) in the laboratory[17]. We detected 662-keV gammas from the point source with a significance of 5$\sigma$ at a distance of 1.5 m with the exposure time of 13 min. The point source increases the dose at the detector front by 0.015 μSv/h from a background dose. If the same amount of gammas entered the ETCC over the whole FoV, the significance would decrease by $5\sigma / \sqrt{100} = 0.5\sigma$, assuming that the background gamma increases proportionally from 1 to 100 to the number of pixels. The current ETCC comprises 100 pixels and one pixel is defined as an area of the unit solid angle in the FoV. In the case of a 100 min observation under the dose of 2 μSv/h at the detector front (assuming the case of Location (C), i.e., low dose), the total number of gammas increases by $0.5\sigma \times \sqrt{(2/0.015) \cdot (100/13)} = 16\sigma$. The expected significance per pixel is then calculated to be $16\sigma / \sqrt{100} = 1.6\sigma$, which is consistent with the observed significances of (1.2 - 2.5$\sigma$) in the low-dose area (see the error bars in Fig.6b). Similarly, the expected significance for the high-dose area is calculated and is found to be also consistent with the observed values of (3 - 5$\sigma$). Thus, our results of the on-site measurements are well consistent with the expected significances estimated from the calibration in the laboratory.

We also estimate the emissivity within the PSF and the detection limit to check consistency with the calibration data. As shown in Fig.7 the covered area by the PSF for the distance L between a target and the ETCC is given by L· sin$\Theta$. Since the number of gammas (brightness) within the PSF is conserved along the line of sight, the sensitivity in the PSF is independent of the distance L if absorption in the air is not taken into account. For example, for the distances L of 10 m and 100 m, the sizes of an area corresponding to a detector pixel are estimated to be 1 m and 10 m, respectively, when the same detection limits for both the distances are used. The detection limit for the ~2$\sigma$ level of the ETCC is 0.5 μSv/h at a unit solid angle for an exposure of 100 min (see the distribution of red points in Fig. 6b). Note that the limit is proportional to $1/\sqrt{(effective\ area \times exposure\ time)}$, and hence can be easily scaled for different exposures and effective areas.

**Acknowledgements**

This study was supported by a Grant in-Aid from the Global COE program "The Next Generation of Physics, Spun from Universality and Emergence" from the MEXT of Japan, and "SENTAN" program promoted by Japan Science and Technology Agency (JST). "SENTAN" program was performed under the leadership of Mr. N. Bando in HORIBA Ltd. with collaboration with Kyoto University and CANON. In particular, we stress that these results could not have been attained without devoted support by Mr. N. Bando, Mr. A. Uesaka, R. Nakamura, T. Watanabe and their colleagues in HORIBA Ltd. for the test measurement with the ETCC in the Fukushima prefecture.

**Author contributions statement**

T.T. is a project leader and wrote this manuscript.

D.T. mainly managed this measurements in Fukushima and analysis, and also wrote the manuscript and made figures.

T.M. mainly contributed to this measurements, analysis and construction of the instrument with writing the manuscript and figures.

A.T. mainly contributed to the design of the instrument and joined the measurement in Fukushima.

S.K. contributed to transfer the basic technology of ETCC to the development of this instrument.

Y.Ma. contributed to develop the data acquisition system of this instrument.

Y.Mi, supported the development of analysis tool and joined the measurement in Fukushima.

M.O. joined the measurement in Fukushima.

**Competing financial interests**

The authors declare no competing financial interests.

**Corresponding Author**

Correspondence and requests for materials should be addressed to T.T. (tanimori@cr.scphys.kyoto-u.ac.jp)

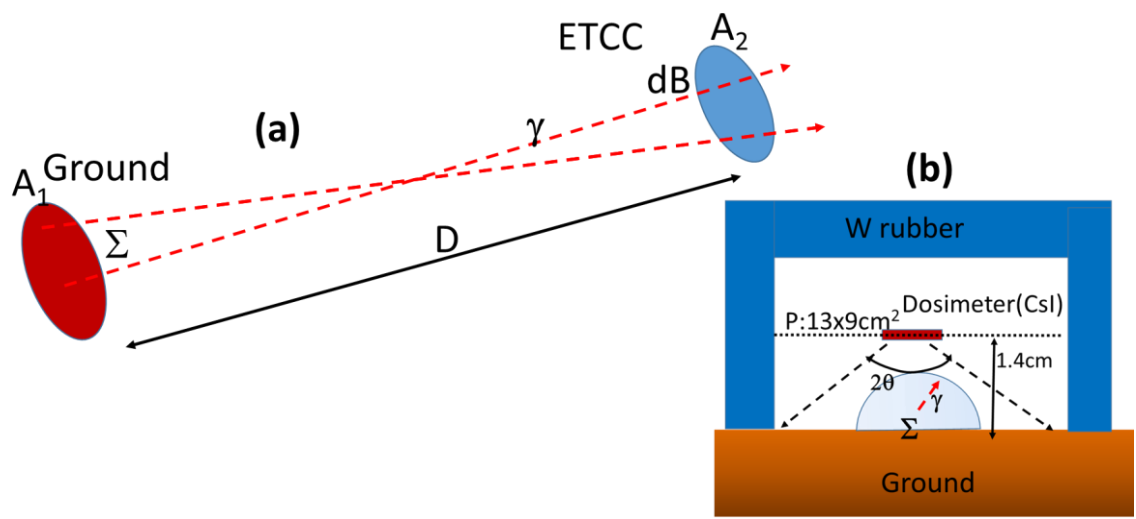

**Figure 1 Schematic explanation of the relation of radioactivity on the ground to the brightness measured by ETCC:**

(a) Schematic explanation of the correlation between emissivity (brightness) (Σ) on the ground and the measured gammas (dB) within a unit solid angle in the FoV of the ETCC, where A1 and A2 indicate the areas on the ground and ETCC, respectively, and D denotes the distance between the ground and the ETCC. (b) Schematic view of the positional relation of the ground, dosimeter (Horiba

PA-1100) and tungsten rubber for our 1-cm dose measurement. The top and four sides of the dosimeter are covered by tungsten rubber to shield it from the downward gamma-ray radiation.

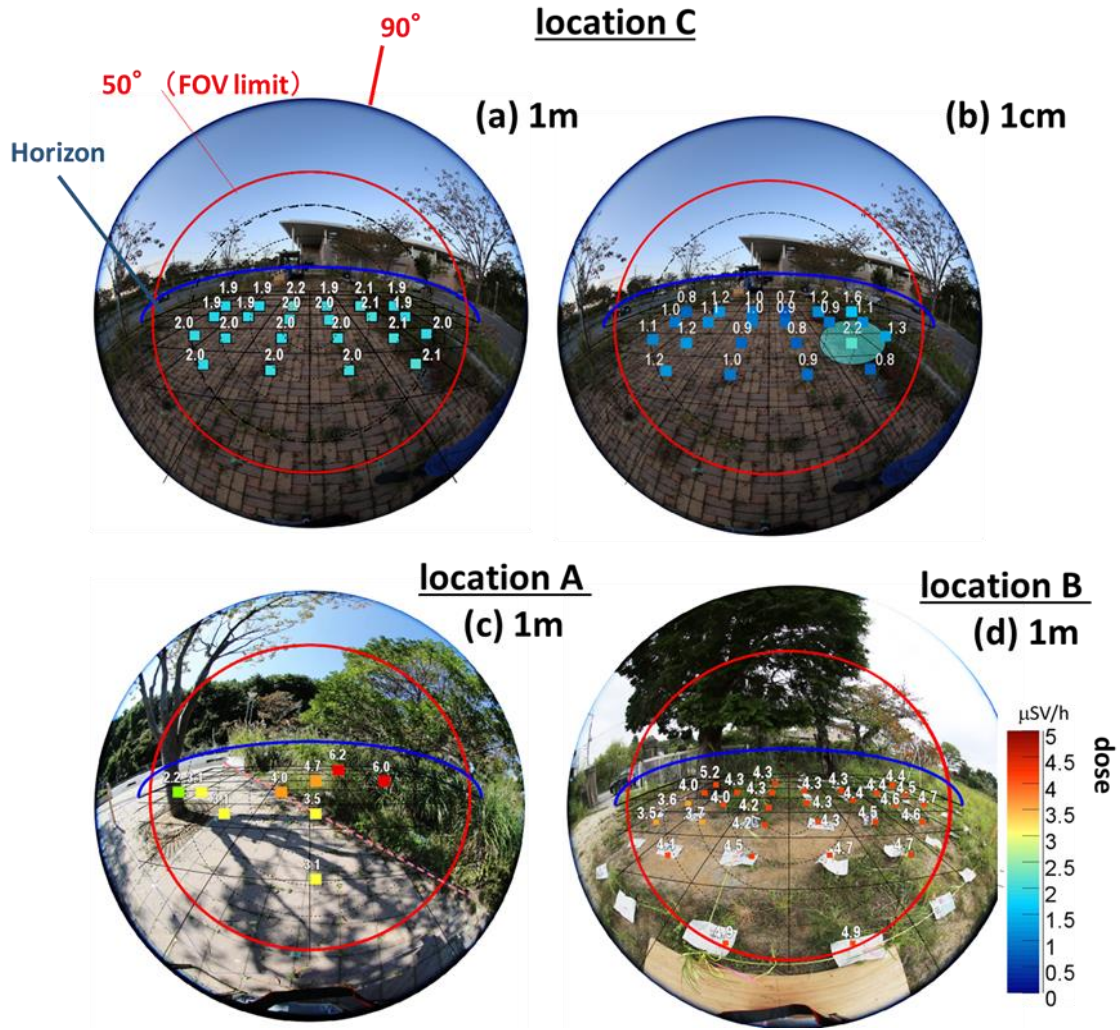

**Figure 2: Comparison between 1-m dose and 1-cm dose**;
Mapping results of (a) 1-m dose and (b) 1-cm dose at a decontaminated parking lot overlaid on the stereo perspective photographs taken with a fish-eye-lens camera. Doses at the height of 1 m and 1 cm on every 1 m square grid in the FoV of the ETCC were mapped as a reference with the commercial dosimeter (HORIBA, Radi PA-1100). See Fig.1b for the setting of the 1-cm dose measurement. Panels (c) and (d) are the mapping results of 1-m dose at the locations (A) and (B), respectively.

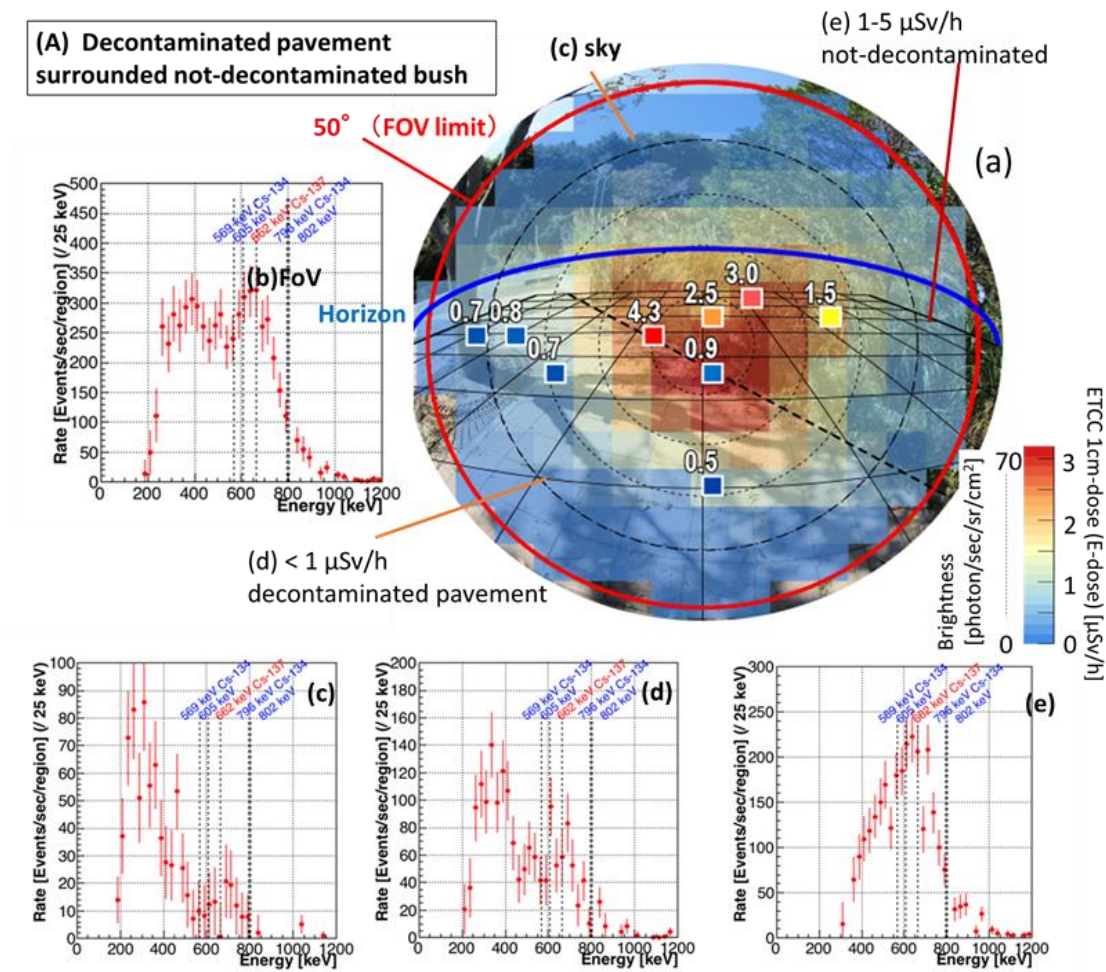

**Figure 3: Energy Spectra at different points in one image of the ETCC (Imaging spectroscopic result):**

Results in the location (A), decontaminated pavement surrounded with not-decontaminated bush. (a) Photograph (as in Fig.2c) overlaid with nine coloured squares for the 1-cm dose and colour map for the E-dose estimated from the ETCC brightness for the energy band of 486-1000 keV. The exposure time is roughly 100 min. The red circle indicates the FoV (100°ϕ). The blue line is the horizon. This mapping is separated into three regions (c, d, and e), from each of which the energy spectrum is accumulated. (b) Observed gamma energy-spectrum for the whole area within the FoV. (c) Sky region (i.e., above the horizon), (d) lowly contaminated region, and (e) highly contaminated region. The spectrum (e) indicates that Cs remains mainly on the surface of the ground and bush, producing the prominent peaks of $^{137}$Cs and $^{134}$Cs. In the spectrum (d), in spite of the decontamination work, Cs remains inside the gaps between tiles on the ground, and still emits line gammas, making up for a half of the brightness of this region, whereas the other half in the spectrum is the scattered gammas coming from Cs on or in the soil. The spectrum (c) shows the similar situation for the sky, where scattered lower-energy gammas appear more clearly than gammas at the Cs peaks.

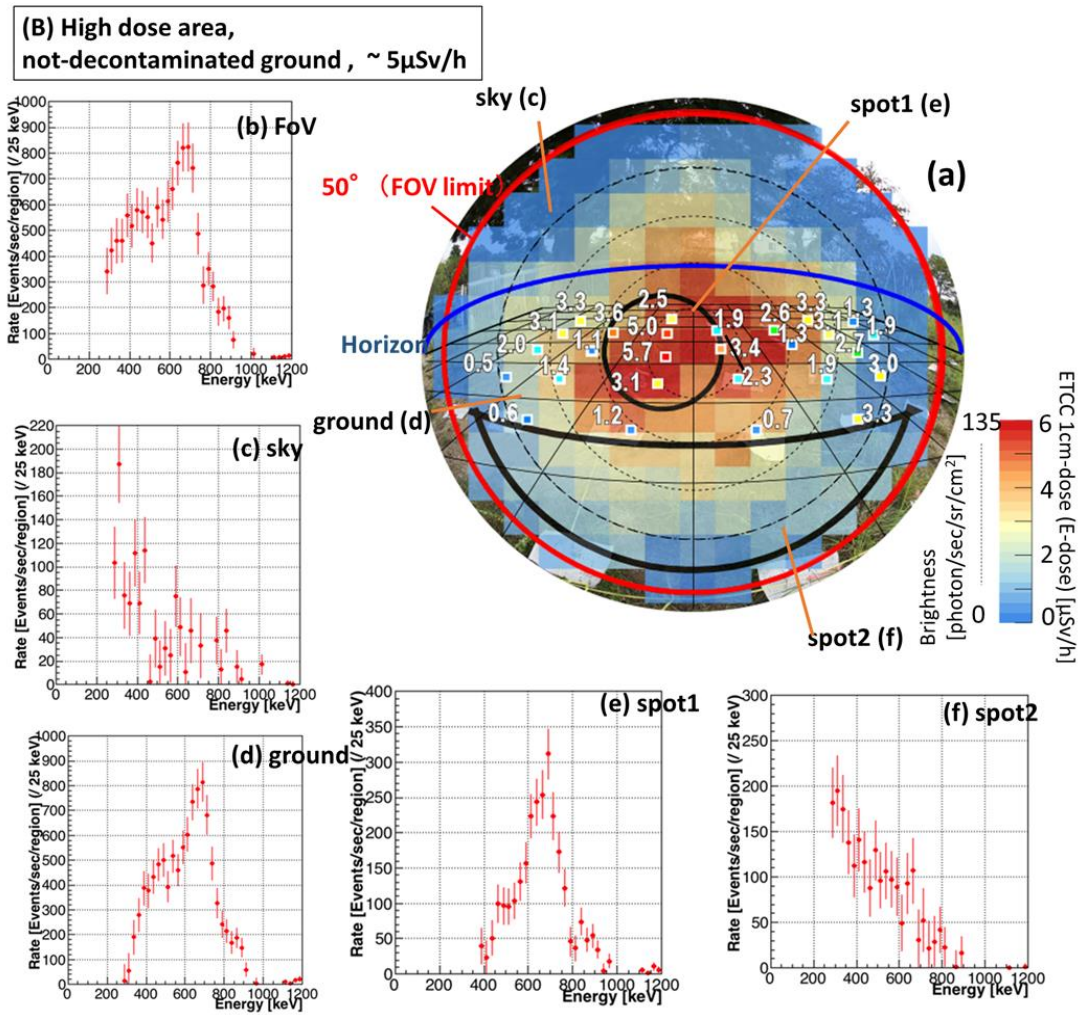

Figure 4: **Imaging Spectroscopic results in contaminated area:**

Results in the location (B): not-decontaminated ground. (a) Photograph (as in Fig.2d) overlaid with the map of 1 cm dose (coloured squares) and the distribution of E-dose (colour image) as in Figure 3a, and five energy spectra of (b) the whole FoV (c) the sky, (d) the ground and (e, f) two sub-parts of the ground, the regions for which are indicated in the panel (a). The exposure time by ETCC is 80 min. Two spots are chosen from the (not-decontaminated) ground region for higher (spot 1, Panel e) and lower (spot 2, Panel f) brightness regions. Their spectra show that the Cs peak is clearly more dominant in the high-dose spot than in the low-dose one.

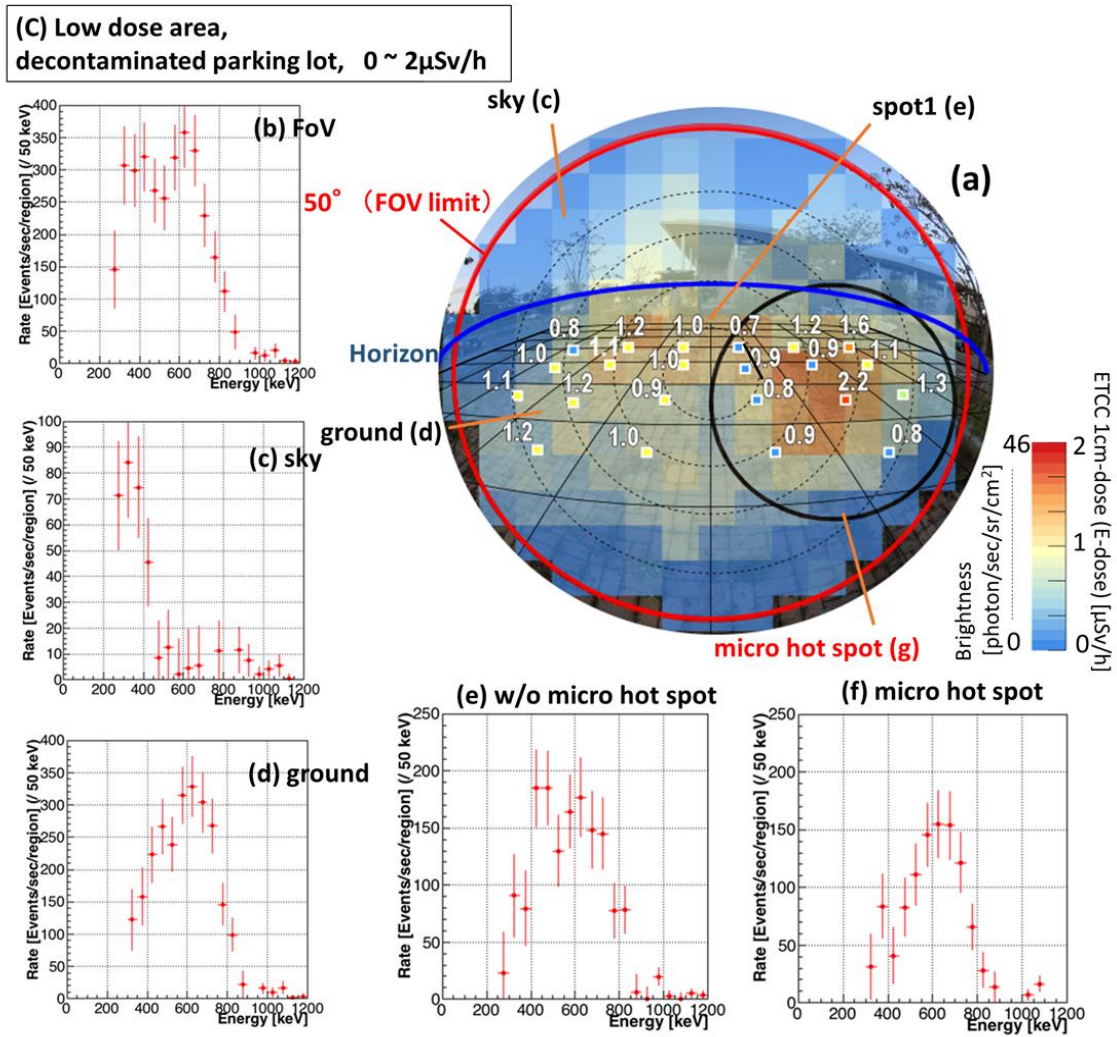

Figure 5: **Imaging Spectroscopic results in decontaminated area:**

Results in the location (C): decontaminated parking lot (Fig.2a). The arrangement of the panels is the same as in Fig.3, except panels (e) and (f), as explained below. The exposure time is 100 min. Both the peaks of Cs and low-energy scattered gammas are seen in (b), whereas a Cs peak in the ground and an associated scattered low-energy tail in the sky dominate the spectra (d) and (c), respectively. The ground is separated into two regions: (f) the micro hot-spot and (e) the rest. The spectra show the prominent Cs peak even in the low-dose micro spot (f), where the dose is only slightly higher than in the surrounding region.

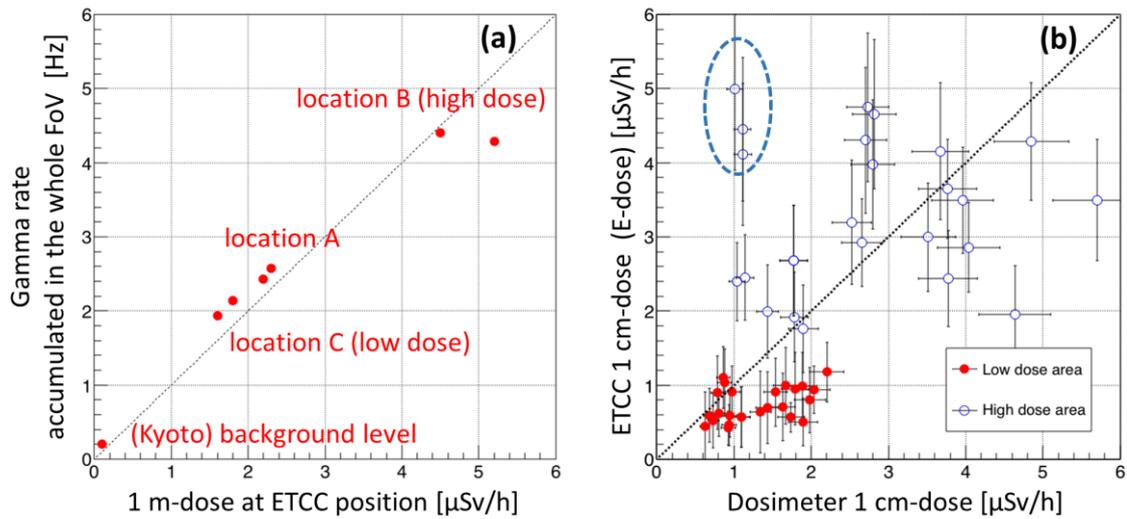

**Figure 6: Correlation plots between the doses measured by a dosimeter and the ETCC:**

(a) Correlation plot between the gamma rate for the whole FoV of the ETCC and 1-m dose at the position of the ETCC.   (b) Correlation plot between the 1-cm dose and E-dose within FoV of 80°ϕ as explained in the caption of Fig.3. The dotted line is the best-fitting result of a linear fit for all the data points except 3 points indicated in a dotted ellipse, of which the doses (or brightness) are strongly affected from the hot spot in Fig.5e due to the proximity of their positions to the hot spot compared with the size of the PSF of the ETCC.

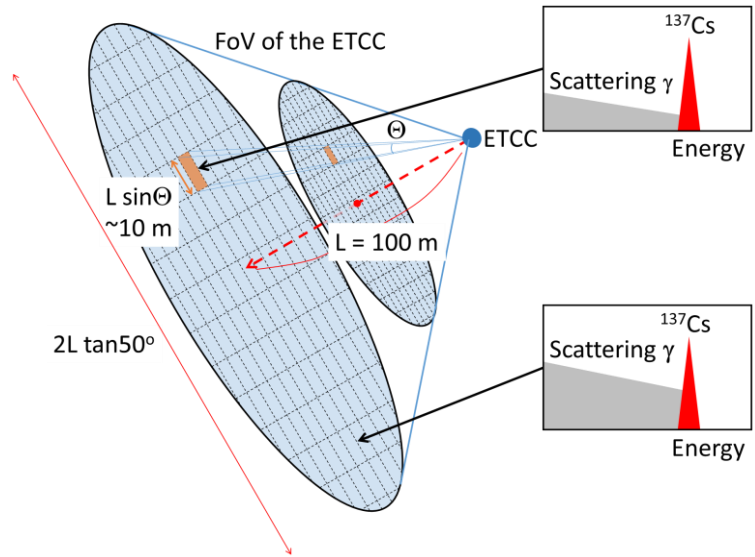

**Figure 7: Schematic explanation of the detection limit of the ETCC from the point of view of geometrical optics.**

Two ellipses filled in light blue show the planes perpendicular to the line of sight. Their areas correspond to the FoV of the ETCC at the respective distances, and are about 1 m x 1 m and 10 m x 10 m at distances of 10 m and 100 m, respectively, from the ETCC. We can perform gamma imaging-spectroscopy as highlighted schematically in the two insets.

Table 1: Averaged 1-m dose by dosimeter, observed brightness measured with the ETCC in the sky and ground, and their ratios, corrected for the difference in the solid angles of sky/ground ~ 1/5.7, in the locations (A), (B), and (C).

|  | (A) Decontaminated pavement and not-decontaminated bush | (B) Not-decontaminated ground | (C) Decontaminated parking lot |
|---|---|---|---|
| **averaged dose [μSv/h]** | 2.2 | 5.2 | 1.8 |
| **gamma from sky [count]** | 776 | 1591 | 432 |
| **gamma from ground [count]** | 4611 | 10163 | 2546 |
| **solid-angle-corrected sky/ground ratio of gamma** | 0.96 | 0.96 | 0.89 |

Table 2: Specifications of current ETCC, ETCC type-A and ETCC type-B. Note that a minimum exposure time is estimated as gammas detected with 2σ significance in each pixel in the condition of 2 μSv/h radiation, which is the same condition as the location (C).

|  | current ETCC | ETCC type-A | ETCC type-B |
|---|---|---|---|
| **TPC volume** | 10 cm x 10 cm x 16 cm | 20 cmφ (in diameter) x 20 cm | 30 cm x 30 cm x 30 cm |
| **TPC gas** | Ar and $C_2H_6$(90 %:10 %) at 1.5 atm | Ar and $CF_4$(50 %:50 %) at 2 atm | $CF_4$ at 3 atm |
| **effective detection area (for 662 keV gammas)** | ~ 0.01 $cm^2$ | ~ 0.2 $cm^2$ | ~ 10 $cm^2$ |
| **minimum exposure time** | ~ 100 min. | several min. | several sec. |